\documentclass[letterpaper, 10 pt, conference]{ieeeconf}  

\usepackage{amsmath, amssymb, amsfonts}
\usepackage[english]{babel}
\usepackage[T1]{fontenc}
\usepackage{epsfig}
\usepackage{array}
\usepackage{subfigure}

\newtheorem{remark}{Remark}

\providecommand{\abs[1]}{\left|#1\right|}

\providecommand{\ve[1]}{\mathbf{#1}}
\newcommand{\guillemets}[1]{``#1''}

\newcommand{\mat}[2]{\left(\begin{array}{#1} #2 \end{array} \right)}

\makeindex

\IEEEoverridecommandlockouts                              
\title{\LARGE \bf Linear identification of nonlinear systems: \\A lifting technique based on the Koopman operator
}
\author{Alexandre Mauroy and Jorge Goncalves
\thanks{A. Mauroy and J. Goncalves are with the Luxembourg Centre for Systems Biomedicine,
        University of Luxembourg, 4367 Belvaux, Luxembourg
        {\tt\small alexandre.mauroy@uni.lu, jorge.goncalves@uni.lu}}%
}

\begin{document}

\maketitle
\thispagestyle{empty}
\pagestyle{empty}

\begin{abstract}

We exploit the key idea that nonlinear system identification is equivalent to linear identification of the so-called Koopman operator. Instead of considering nonlinear system identification in the state space, we obtain a novel \emph{linear} identification technique by recasting the problem in the infinite-dimensional space of observables. This technique can be described in two main steps. In the first step, similar to a component of the Extended Dynamic Mode Decomposition algorithm, the data are lifted to the infinite-dimensional space and used for linear identification of the Koopman operator. In the second step, the obtained Koopman operator is \guillemets{projected back} to the finite-dimensional state space, and identified to the nonlinear vector field through a linear least squares problem. The proposed technique is efficient to recover (polynomial) vector fields of different classes of systems, including unstable, chaotic, and open systems. In addition, it is robust to noise, well-suited to model low sampling rate datasets, and able to infer network topology and dynamics.
\end{abstract}

\section{Introduction}

Operator-theoretic techniques rely on the idea of lifting nonlinear dynamical systems to an infinite dimensional space, where those systems have a linear representation. For instance, the so-called Koopman operator is a linear operator that describes the evolution of observable-functions along the trajectories of the system. Its spectral properties have been investigated over the past years and related to geometric properties of the system (see e.g. \cite{Mezic}). On the one hand, this body of work yielded novel methods for the analysis of nonlinear systems described by a known vector field (e.g. global stability analysis \cite{Mauroy_Mezic_stability}, global linearization \cite{Lan}, analysis of monotone systems \cite{Sootla_Mauroy}). On the other hand, the Koopman operator approach was shown to be conducive to data analysis \cite{Rowley} and related to numerical methods such as Dynamic Mode Decomposition (DMD) \cite{Schmid}, yielding new techniques for the analysis of nonlinear systems described by trajectory data points (snapshots). In this context, the present paper aims at bridging these two directions of analysis, connecting data to vector field.

While Koopman operator techniques have been previously considered in the framework of system analysis, we present a first attempt --to the authors knowledge-- to develop them for system identification. Exploiting the lifting obtained through the Koopman operator, our key idea is to \emph{identify a linear operator in the space of observables}, instead of identifying a nonlinear system in the state space. Since Koopman operator and state space representations are both equivalent, identifying the operator is indeed sufficient to recover the nonlinear dynamical system. Because the Koopman operator is defined in an infinite-dimensional space, we consider its projection onto a finite basis of functions. The proposed method can somehow be interpreted, in the space of observables, as an analog of methods developed in \cite{Brunton, Wei}, which use dictionary functions in the state space. Our method proceeds in two main steps: from state space to space of observables, and back to state space. In the first step, a finite-dimensional projection of the Koopman operator is identified from data through classical linear identification, a method that is related to a component of the so-called Extended Dynamic Mode Decomposition (EDMD) algorithm \cite{Rowley_EDMD}. In the second step, the projected operator is connected to the vector field of the nonlinear system through an analytic derivation of the infinitesimal generator (following similar steps as in \cite{Mauroy_Mezic_stability}). The expansion of the vector field in the finite basis is then obtained by solving a linear least squares problem.

Thanks to the lifting technique, the proposed method has the advantage of relying only on linear techniques (i.e. least squares regression), in spite of the fact that the dynamical system is nonlinear. In contrast to recent (direct) methods (e.g. \cite{Brunton}), it is a (indirect) method that does not rely on the estimation of time derivatives, which cannot be performed when the sampling rate is too low. Therefore, the method is well-suited to low sampling, and also robust to both measurement noise and process noise. In addition, it works efficiently with different kinds of behaviors, including unstable and chaotic systems (although limited so far to polynomial vector fields) and can be applied to several small datasets obtained from multiple trajectories. This flexibility is well-suited to identification problems such as those encountered in the context of system biology.

The rest of the paper is organized as follows. In Section \ref{sec:Koopman}, we introduce the system identification problem and present the lifting technique obtained through the Koopman operator. Our numerical method for nonlinear system identification based on the lifting technique is described in Section \ref{sec:num}. Extensions of the method to open systems (i.e. with outputs or process noise) are discussed in Section \ref{sec:exten}. In Section \ref{sec:examples}, the method is illustrated with several numerical examples, and applied to network reconstruction. Concluding remarks and perspectives are given in Section \ref{conclu}.
\newpage

\section{Koopman operator technique \\for system identification}
\label{sec:Koopman}

\subsection{Problem statement}

Consider a (unknown) dynamical system
\begin{equation}
\label{syst1}
\dot{\ve{x}} = \ve{F}(\ve{x}) \,, \quad \ve{x}\in \mathbb{R}^n
\end{equation}
with a vector field $\ve{F}(\ve{x})$ that is assumed to be polynomial. We denote by $\varphi:\mathbb{R}^+ \times \mathbb{R}^n \to \mathbb{R}^n$ the flow induced by the system, i.e. $t \mapsto \varphi^t(\ve{x_0})=\varphi(t,\ve{x_0})$ is a solution of \eqref{syst1} associated with the initial condition $\ve{x_0}$. 

We address the problem of identifying the vector field $\ve{F}$ given $K$ snapshot pairs $(\ve{x}_k,\ve{y}_k) \in \mathbb{R}^{n \times 2}$. The data points are such that $\ve{y}_k=\varphi^{T_s}(\ve{x}_k)(\ve{1}+\ve{v}_k)$, where each component of the measurement noise $\ve{v}_k$ is a Gaussian random variable with zero mean and standard deviation $\sigma_{meas}$. The sampling period $T_s$ is assumed to be identical for all pairs $(\ve{x}_k,\ve{y}_k)$. Note that the data points $\ve{x}_k$ and $\ve{y}_k$ can belong to different trajectories. Stochastic systems with process noise and systems with inputs will also be considered in Section \ref{sec:exten}.

\subsection{Lifting technique using the Koopman operator}

The system \eqref{syst1} is described by its flow $\varphi^t$ in the state space $\mathbb{R}^n$. Alternatively, the system can be lifted to an infinite-dimensional space $\mathcal{F}$ of observable-functions $f:\mathbb{R}^n \to \mathbb{R}$. In this space, it is described by the semigroup of Koopman operators $U^t:\mathcal{F} \to \mathcal{F}$, $t\geq 0$, which governs the evolution of the observables along the trajectories:
\begin{equation}
\label{def_Koopman}
U^t f=f \circ \varphi^t \quad \forall f \in \mathcal{F}\,.
\end{equation}
This lifting technique is of interest since it leads to a linear (but infinite-dimensional) representation of the system. Even if the system \eqref{syst1} is nonlinear, the operator is always linear since we have
\begin{equation*}
U^t (a_1 f_1 + a_2 f_2) = a_1 f_1 \circ \varphi^t + a_2 f_2 \circ \varphi^t = a_1 U^t f_1 + a_2 U^t f_2\,.
\end{equation*}
In addition, there is a bijective relationship between the Koopman operator and its associated system. We denote by $L=\lim_{t \rightarrow 0} (U^t-I)/t$ the infinitesimal generator of the Koopman semigroup, which also satisfies
\begin{equation}
\label{Ut_L}
U^t = e^{L t } = \sum_{k=0}^\infty \frac{t^k}{k!} L^k\,.
\end{equation}
Provided that the vector field $\ve{F}$ is continuously differentiable, it can be shown that
\begin{equation}
\label{inf_gen}
L = \ve{F} \cdot \nabla\,,
\end{equation}
where $\nabla$ denotes the gradient (see e.g. \cite{Lasota_book}). This implies that there is a one-to-one correspondence between the Koopman operator and the vector field.

\begin{figure}[h]
   \centering
    \includegraphics[width=7cm]{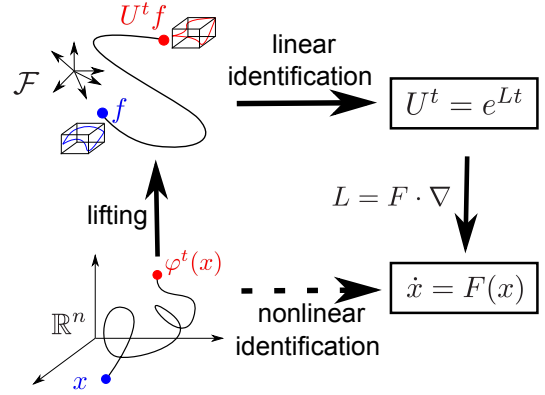}
   \caption{Classical nonlinear system identification is performed directly in the state space. In contrast, the proposed Koopman operator identification technique consists of two steps: (1) lifting of the data and linear identification of the Koopman operator; (2) identification of the vector field using \eqref{Ut_L} and \eqref{inf_gen}.}
   \label{Koopman_ident}
\end{figure}

This paper exploits the equivalence between the two descriptions of the system, proposing to identify the linear Koopman operator in the space of observables. Figure \ref{Koopman_ident} illustrates the two main steps of the method. In the first step, data are lifted to the space of observables, providing the evolution of observables (at some points in the state space). Then classical linear identification is used to obtain an approximation of the Koopman operator (in a finite-dimensional subspace of $\mathcal{F})$. This corresponds to a component of the so-called Extended Dynamic Mode Decomposition (EDMD) algorithm \cite{Rowley_EDMD}. In the second step, the vector field $\ve{F}$ is obtained through the equalities \eqref{Ut_L} and \eqref{inf_gen}.

\section{Numerical method}
\label{sec:num}

This section describes the numerical scheme derived from the Koopman operator identification method. Lifting of the data and identification of the operator (step 1) are explained in the first subsection. Identification of the vector field (step 2) is described in the second subsection. 

Conceptually, the method works in the infinite-dimensional space $\mathcal{F}$ of observables. But to be numerically tractable, it is developed on a finite-dimensional subspace $\mathcal{F}_N$ which is spanned by a basis of $N$ linearly independent functions $\{p_k\}_{k=1}^N$. In our case, we will choose a basis of monomials with total degree less or equal to $m$:
\begin{equation*}
p_k(\ve{x}) \in \{ x_1^{s_1} \cdots x_n^{s_n} | (s_1,\dots,s_n) \in \mathbb{N}^n, s_1+\cdots+s_n \leq  m \} \,.
\end{equation*}
This choice is motivated by the fact that the vector field is assumed to be polynomial. Other bases of functions (e.g. Fourier basis, radial basis functions) could be considered to generalize the method. For practical purposes, the sequence of monomials should be characterized by some order (e.g. lexicographic order, weight order). The number of functions in the basis is equal to $N=(n+m)!/(n! \, m!)$. We denote by $\ve{p}(\ve{x}) = (p_1(\ve{x}),\dots,p_N(\ve{x}))^T$ the vector of basis monomials.

The Koopman operator $U^t$ can be projected onto the subspace $\mathcal{F}_N$ and represented by a matrix $\ve{\overline{U}}$. Considering
\begin{equation}
\label{f_expan}
f(\ve{x})= \ve{a}^T \ve{p}(\ve{x}) \in \mathcal{F}_N
\end{equation}
and 
\begin{equation}
\label{Uf_expan}
\mathbb{P} U^t f(\ve{x}) = \ve{b}^T \ve{p}(\ve{x}) \in \mathcal{F}_N
\end{equation}
where $\mathbb{P}:\mathcal{F} \to \mathcal{F}_N$ is a projection operator, we have
\begin{equation}
\label{def_U_bar}
\ve{\overline{U}} \, \ve{a} = \ve{b}\,, \qquad \ve{\overline{U}} \in \mathbb{R}^{N \times N} \,.
\end{equation}
The matrix $\ve{\overline{U}}$ provides a (approximate) finite-dimensional linear representation of the nonlinear system. This representation is not obtained through local linearization techniques and is valid globally in the state space. In the numerical scheme described below, we compute $\ve{\overline{U}}$ from data (step 1), and then connect the obtained matrix to the vector field of the nonlinear system (step 2).

\subsection{Step 1: Lifting of the data and Koopman operator identification}
\label{subsec_step1}

We now describe the method used to compute the matrix $\ve{\overline{U}}$ given snapshot pairs $\{(\ve{x}_k,\ve{y}_k)\}_{k=1}^K$, which corresponds to a component of the EDMD algorithm (see \cite{Rowley_EDMD} for more details).

For each snapshot pair $(\ve{x}_k,\ve{y}_k) \in \mathbb{R}^{n  \times 2}$, we construct the new pair $(\ve{p(x_k)},\ve{p(y_k)}) \in \mathbb{R}^{N  \times 2}$. The data are somehow lifted to (a subspace of) the space of observables and provide snapshots of trajectories of particular observables (i.e. the basis functions). Provided that the measurement noise $\ve{v}_k$ is small, we have $\ve{y}_k \approx \varphi^{T_s}(\ve{x}_k)$. It follows from \eqref{def_Koopman} and \eqref{f_expan} that
\begin{equation*}
\mathbb{P} U^{T_s} f(\ve{x}_k) = \ve{a}^T \mathbb{P} U^{T_s} \ve{p(\ve{x}_k)} \approx \ve{a}^T \mathbb{P} \ve{p(\ve{y}_k)}
\end{equation*}
and from \eqref{Uf_expan} and \eqref{def_U_bar} that
\begin{equation*}
\mathbb{P} U^{T_s} f(\ve{x}_k) = \ve{a}^T \ve{\overline{U}}^T \ve{p}(\ve{x}_k)\,,
\end{equation*}
so that
\begin{equation}
\label{rel_py_px}
\ve{p}(\ve{x}_k)^T \, \ve{\overline{U}} \approx \mathbb{P} \ve{p}(\ve{y}_k)^T
\end{equation}
where $\ve{\overline{U}}$ denotes here the matrix representation of $\mathbb{P} U^{T_s}$. 
 The last equality implies that the $j$th column $\ve{c}_j$ of $\ve{\overline{U}}$ is so that $\ve{c}_j^T \ve{p} \approx \mathbb{P} U^{T_s} p_j$. If the projection $\mathbb{P}$ is chosen so that it yields the least squares fit at the points $\ve{x}_k$, $ k=1,\dots,K$ (i.e. finite-dimensional Galerkin projection), the matrix $\ve{\overline{U}}$ is given by the least squares solution
\begin{equation}
\label{U_data}
\ve{\overline{U}} = (\ve{P_x})^\dagger \, \ve{P_y}
\end{equation}
with the $K \times N$ matrices
\begin{equation}
\label{mat_P}
\ve{P_x} = \mat{c}{\ve{p}(\ve{x}_1)^T \\ \vdots \\ \ve{p}(\ve{x}_K)^T} 
\qquad 
\ve{P_y} = \mat{c}{\ve{p}(\ve{y}_1)^T \\ \vdots \\ \ve{p}(\ve{y}_K)^T}
\end{equation}
and where $\ve{P}^\dagger$ denotes the pseudoinverse of $\ve{P}$. 

\subsection{Step 2: Identification of the vector field}
\label{subsec_step2}

We assume that the polynomial vector field $\ve{F}=(F_1,\dots,F_n)^T$ is of total degree less or equal to $m_F$, so that we can write
\begin{equation}
\label{expan_F}
F_j(\ve{x}) = \sum_{k=1}^{N_F} w^{j}_k p_k(\ve{x}) \qquad j=1,\dots,n
\end{equation}
where $N_F=(m_F+n)!/(m_F! n!)$ is the number of monomials $p_k$ of total degree less or equal to $m_F$. The identification problem is therefore reduced to identifying $n N_F$ coefficients $w^{j}_k$.

\subsubsection{Derivation of the Koopman infinitesimal generator} Our goal is to derive the matrix $\ve{\overline{L}}$ which represents the projection of the infinitesimal generator \eqref{inf_gen} on the basis of monomials $p_k$, and show that this is a function of the unknown coefficients $w^{j}_k$ of the vector field. Consider the linear operators
\begin{equation*}
L^{j}_k = p_k \, \frac{\partial}{\partial x_j}\,, \quad j=1,\dots,n\,, \quad k=1,\dots,N_F \,.
\end{equation*}
It is clear that the operators $L_k^j$ map polynomials of total degree less or equal to $m_1$ to polynomials of total degree less or equal to $m_2=m_1+m_F-1$. Denote by $\ve{p}^{m}(\ve{x})$ the vector of monomials of total degree less or equal to $m$. The projection of $L^{j}_k$ on the basis of monomials is represented by the matrix $\ve{\overline{L}}_k^j \in \mathbb{R}^{N_2\times N_1}$, with $N_1=(m_1+n)!/(m_1! n!) \leq N_2=(m_2+n)!/(m_2! n!)$, which is so that
\begin{equation}
\label{f_L_jk_bis}
L^j_k f(\ve{x}) = (\ve{\overline{L}}_k^j \ve{a})^T \ve{p}^{m_2}(\ve{x})
\end{equation}
for all $f(\ve{x})=\ve{a}^T \ve{p}^{m_1}(\ve{x})$. We have also $L_k^j \ve{p}^{m_1}(\ve{x}) = (\ve{\overline{L}}^j_k)^T \ve{p}^{m_2}(\ve{x})$, which implies that the $l$th column of $\ve{\overline{L}}^j_k$ corresponds to the expansion of $L_k^j p_l$ in the basis of monomials $\ve{p}^{m_2}$. Defining an index function $\Psi=(\psi_1(k),\dots,\psi_n(k))$ that encodes the order of the monomials in the vector $\ve{p}$, i.e. $p_k(\ve{x})=x_1^{\psi_1(k)} \cdots x_n^{\psi_n(k)}$, we have
\begin{equation*}
L_k^j p_l = \psi_j(l) p_{\Psi^{-1}(\Psi(k)+\Psi(l)-\ve{e}_j)}
\end{equation*}
where $\ve{e}_j \in \mathbb{R}^n$ is the $j$th unit vector. It follows that the entries of $\ve{\overline{L}}^j_k$ are given by
\begin{equation}
\label{L_jk}
\left[\ve{\overline{L}}^j_k \right]_{il} =
\begin{cases} 
\psi_j(l) & \textrm{if } \Psi(i)=\Psi(k)+\Psi(l)-\ve{e}_j \,,\\
0 & \textrm{otherwise}\,.
\end{cases}
\end{equation}
Note that the matrices $\ve{\overline{L}}^j_k$ can also be obtained by multiplying a multiplication matrix and a differentiation matrix (see \cite{Mauroy_Mezic_stability} for more details). Finally it follows from \eqref{inf_gen} and \eqref{expan_F} that
\begin{equation}
\label{L_sum}
L = \sum_{j=1}^n \sum_{k=1}^{N_F} w^j_k \, L^j_k
\end{equation}
and the matrix $\ve{\overline{L}} \in \mathbb{R}^{N_2 \times N_1}$ given by
\begin{equation}
\label{L_w_jk}
\ve{\overline{L}} = \sum_{j=1}^n \sum_{k=1}^{N_F} w^j_k \, \ve{\overline{L}}^j_k
\end{equation}
represents the projection of the Koopman infinitesimal generator on the basis of monomials $p_k$.

\subsubsection{Computation of the vector field} The coefficients $w_k^j$ of the vector field are finally computed by using the relationship between $\ve{\overline{U}}$ (obtained from data in \eqref{U_data}) and $\ve{\overline{L}}$ (which depends on $w_k^j$ in \eqref{L_w_jk}).

It can be shown easily (using \eqref{Ut_L}) that, if $\ve{\overline{U}}$ and $\ve{\overline{L}}$ represent the projection of $U^t$ and $L$, respectively, then $\ve{\overline{U}} = e^{\ve{\overline{L}} t}$.
This property and \eqref{U_data} imply that we can obtain an estimation of $\ve{\overline{L}}$ from data:
\begin{equation}
\label{L_data}
\ve{\overline{L}}_{data} = \frac{1}{T_s} \log((\ve{P_x})^\dagger \, \ve{P_y}) \in \mathbb{R}^{N \times N}\,.
\end{equation} 
The function $\log$ denotes the (principal) matrix logarithm.
\begin{remark}[Matrix logarithm]
The real eigenvalues of $\ve{\overline{U}}=(\ve{P_x})^\dagger \, \ve{P_y}$ shall not be negative, so that the principal logarithm exists and is real. However, numerical simulations show that $\ve{\overline{U}}$ can have negative real eigenvalues when the number of data points is too small. In this case, there exists no real matrix $\log(\ve{\overline{U}})$ and the method might fail.

When the Koopman infinitesimal generator admits complex eigenvalues, the sampling period $T_s$ must be sufficiently small so that the eigenvalues of $T_s \ve{\overline{L}}$ lie in the strip $\{z\in \mathbb{C}:|\Im\{z\}|<\pi\}$. If this condition is not satisfied, $\log(\exp(T_s \ve{\overline{L}})) \neq T_s \ve{\overline{L}}$ and the accuracy of the method is altered. This issue is discussed with more details in \cite{Zuogong_CDC}. \hfill $\diamond$
\end{remark}

Using \eqref{L_w_jk}, we obtain the equality
\begin{equation}
\label{equality_final}
\ve{\overline{L}} = \sum_{j=1}^n \sum_{k=1}^{N_F} w^j_k \, \ve{\overline{L}}^j_k = \ve{\hat{\overline{L}}}_{data}
\end{equation}
where $\ve{\hat{\overline{L}}}_{data}$ is the $N_2 \times N_1$ matrix constructed with the $N_1$ columns and $N_2$ rows of $\ve{\overline{L}}_{data}$ associated with monomials of total degree less or equal to $m_1$ and $m_2$, respectively (note that we will typically choose $N=N_2 \geq N_1$ for the size of $\ve{\overline{U}}$ and $\ve{\overline{L}}_{data}$). By disregarding the $N_2-N_1$ remaining columns of $\ve{\overline{L}}_{data}$, we only consider monomials that are mapped by $L$ onto the span of basis monomials (of total degree less or equal to $m_2$), for which the finite Galerkin projection (i.e. the identity in this case) is exact.

The matrix equality \eqref{equality_final} corresponds to a linear set of equations (with $n N_F$ unknowns $w_k^j$) that is overdetermined. Its least squares solution yields an approximation of the coefficients $w_k^j$:
\begin{equation}
\label{equality_final2}
\mat{c}{w_1^1 \\ \vdots \\ w_{N_F}^n} = \mat{ccc}{| & & | \\ \mathbf{vec} (\ve{\overline{L}}^1_1) & \dots &  \mathbf{vec} (\ve{\overline{L}}^n_{N_F}) \\ | & & |}^\dagger \mathbf{vec}(\ve{\hat{\overline{L}}}_{data})
\end{equation}
where $\mathbf{vec}$ stands for the vectorization of the matrix. Note that more advanced techniques could be used to solve \eqref{equality_final}, for instance promoting sparsity of the vector of unknowns $w_k^j$ by adding a $L^1$-norm penalty term to \eqref{equality_final}, see e.g. the LASSO method \cite{Lasso}. We leave this for future research.

\begin{remark} Each entry of the matrix $\ve{\overline{L}}$ yields an equality in \eqref{equality_final2}. However, if a given entry of $\ve{\overline{L}}^j_k$ is zero for all $j$ and $k$, then the corresponding entry of $\ve{\overline{L}}$ does not depend on the coefficients $w_k^j$ and the related equality in \eqref{equality_final2} can be disregarded. In particular, when $m_1=1$, the number of effective equalities is equal to $n N_F$ and the problem is not overdetermined. Moreover, \eqref{equality_final} can be solved directly in this case since each coefficient $w_k^j$ is equal to an entry of $\ve{\overline{L}}$.
\hfill $\diamond$
\end{remark}
\begin{remark}
The identification problem could also be performed at the level of the Koopman semigroup. However solving the equality $\ve{\overline{U}} = e^{\ve{\overline{L}}T_s}$ (with a square matrix $\ve{\overline{L}}$) amounts to solving a (nonconvex) nonlinear least squares problem. Numerical simulations suggest that better results are obtained by solving \eqref{equality_final}.
\hfill $\diamond$
\end{remark}

%

\section{Extensions to open systems}
\label{sec:exten}

In this section, we show that the proposed method is well-suited to identify open systems that are driven by a known input or by a white noise (i.e. process noise).

\subsection{Systems with inputs}

Consider an open dynamical system of the form
\begin{equation}
\label{input_syst}
\dot{\ve{x}} = \ve{F}(\ve{x},\ve{u}(t))
\end{equation}
with $\ve{x} \in \mathbb{R}^n$ and with the input $u \in \mathcal{U}:\mathbb{R}^+ \to \mathbb{R}^p$. We define the associated flow $\varphi:\mathbb{R}^+ \times \mathbb{R}^n \times \mathcal{U}$ so that $t \mapsto \varphi(t,\ve{x},\ve{u(\cdot)})$ is a solution of \eqref{input_syst} with the initial condition $\ve{x}$ and the input $\ve{u(\cdot)}$. Following the generalization proposed in \cite{Proctor_input} for discrete-time dynamical systems, we consider observables $f:\mathbb{R}^n \times \mathbb{R}^p \to \mathbb{R}$ and define the semigroup of Koopman operators
\begin{equation*}
U^t f(\ve{x,u}) = f(\varphi^t(\ve{x,u(\cdot) = \ve{u}}),\ve{u})
\end{equation*}
where $\ve{u(\cdot)} = \ve{u}$ is a constant input. In this case, $\ve{u}$ can be considered as additional state variables and the above operator is the classical Koopman operator for the augmented system $\dot{\ve{x}}=\ve{F(x,u)}$, $\dot{\ve{u}} = \ve{0}$. In particular, the infinitesimal generator is still given by \eqref{inf_gen}.

It follows that the method proposed in Sections \ref{subsec_step1} and \ref{subsec_step2} can be used if the input can be considered as constant between two snapshots (zero-order hold assumption), or equivalently if the sampling rate is high enough. The matrix $\ve{\overline{U}}$ is obtained with snapshot pairs $([\ve{x}_k,\ve{u}_k],[\ve{y}_k,\ve{u}_k]) \in \mathbb{R}^{(n+p) \times 2}$ and the rest of the procedure follows on similar lines with the augmented state space $\mathbb{R}^{n+p}$. The efficiency of the identification method in the case of systems with inputs is shown in Section \ref{sec:example_input}.

\subsection{Process noise}

We have considered so far only measurement noise. We will show that the proposed method is also robust to process noise. Consider a system described by the stochastic differential equation
\begin{equation}
\label{syst_stoch}
\dot{x}_k = F_k(\ve{x}) + \eta_k(t)
\end{equation}
where $\eta_k(t)$ is a white noise that satisfies $\mathbb{E}[\eta_k(s) \eta_j(t)]= \sigma^2_{proc} \delta_{kj} \delta(t-s)$ (where $\mathbb{E}$ denotes the expectation). We define the flow $\varphi:\mathbb{R}^+ \times \mathbb{R}^n \times \Omega$, where $\Omega$ is the probability space, such that $t \mapsto \varphi(t,\ve{x},\omega)$ is a solution of \eqref{syst_stoch}. In this case, the semigroup of Koopman operators is defined by (see e.g. \cite{Mezic})
\begin{equation*}
U^t f (\ve{x}) = \mathbb{E}[f(\varphi(t,\ve{x},\omega))]
\end{equation*}
and its infinitesimal generator is given by
\begin{equation*}
L f = \ve{F} \cdot \nabla f + \frac{\sigma^2_{proc}}{2} \Delta f
\end{equation*}
where $\Delta=\sum_k \partial^2/\partial x_k^2$ denotes the Laplacian operator that accounts for diffusion. Note that the infinitesimal generator is related to the so-called Kolmogorov backward equation.

Now we can show that the numerical scheme of the proposed identification method does not need to be adapted to take account of process noise. As explained in \cite{Rowley_EDMD}, the first step of the method (Section \ref{subsec_step1}) is still valid for identifying the matrix $\ve{\overline{U}}$. In the second step (Section \ref{subsec_step2}), the procedure is the same, except that one has to consider the Laplacian operator whose projection on the basis of monomials is represented by a matrix $\ve{\overline{D}} \in \mathbb{R}^{N_2 \times N_1}$. The equality \eqref{equality_final} is then replaced by
\begin{equation*}
 \sum_{j=1}^n \sum_{k=1}^{N_F} w^j_k \, \ve{\overline{L}}^j_k + \frac{\sigma^2}{2} \ve{\overline{D}} = \ve{\hat{\overline{L}}}_{data}
\end{equation*}
 where $\sigma$ is an additional unknown. While the operators $L^j_k$ map monomials of total degree $m$ to monomials of total degree greater or equal to $m-1$, the Laplacian operator maps monomials of total degree $m$ to monomials of total degree $m-2$. Therefore all nonzero entries of $\ve{\overline{D}}$ correspond to zero entries of $\ve{\overline{L}}^j_k$, so that the addition of the diffusion term only modifies entries of $\ve{\overline{L}}$ which are zero---and do not depend on $w_k^j$---when there is no process noise. In other words, the diffusion term does not affect the equalities on $w_k^j$, whose solution is still given by \eqref{equality_final2}. In Section \ref{sec:example_input}, an example illustrates the robustness of the method against process noise.

\section{Numerical examples}
\label{sec:examples}

In this section, we illustrate our numerical method with several examples. We show that the method is efficient to recover the vector field of different kinds of systems, including unstable, chaotic, and open systems. The method is also applied to network reconstruction.

For all the simulations, we consider measurement noise with standard deviation $\sigma_{meas}=0.01$. We also choose the parameters $m_1=1$ and $m_F=3$, so that $m_2=m_1+m_F-1=3$.

\subsection{Periodic, unstable, and chaotic systems}

We first consider three systems that exhibit different kinds of behavior.

\paragraph{Van der Pol oscillator} The dynamics are given by
\begin{eqnarray*}
\dot{x}_1 & = & x_2 \\
\dot{x}_2 & = & (1-x_1^2)x_2-x_2
\end{eqnarray*}
and possess a stable limit cycle. The data are obtained by taking $3$ snapshots at $t\in\{0,0.5,1\}$ (sampling period $T_s=0.5$) for $10$ trajectories with initial conditions on $[-1,1]^2$.

\paragraph{Unstable equilibrium} The dynamics are given by
\begin{eqnarray*}
\dot{x}_1 & = & 3 \, x_1+0.5 \, x_2 -x_1 x_2 + x_2^2 + 2 \, x_1^3 \\
\dot{x}_2 & = & 0.5 \,x_1+4 \, x_2
\end{eqnarray*}
and are characterized by an unstable equilibrium at the origin. The data are obtained by taking $2$ snapshots at $t \in\{0,0.1\}$ (sampling period $T_s=0.1$) for $20$ trajectories with initial conditions on $[-1,1]^2$.

\paragraph{Chaotic Lorenz system} The dynamics are given by
\begin{eqnarray*}
\dot{x}_1 & = & 10(x_2-x_1) \\
\dot{x}_2 & = & x_1(28-x_3)-x_2 \\
\dot{x}_3 & = & x_1 x_2 - 8/3 \, x_3
\end{eqnarray*}
and exhibit a chaotic behavior. The data are obtained by taking $31$ snapshots over $[0,1]$ (sampling period $T_s=0.033$) for $10$ trajectories with initial conditions on $[-20,20]^3$.

For each model, we identify the coefficient $w_k^j$ of the vector field and compute the root mean square error
\begin{equation*}
\textrm{RMSE} = \sqrt{\frac{1}{n N_F} \sum_{j=1}^n \sum_{k=1}^{N_F} \left((w_k^j)_{estim} - (w_k^j)_{exact} \right)^2}\,. 
\end{equation*}
where $(w_k^j)_{estim}$ and $(w_k^j)_{exact}$ are the estimated and exact values of the coefficients $w_k^j$. In order to compare the results obtained with different systems, we also compute the normalized RMSE: $\textrm{NRMSE} = \textrm{RMSE}/\overline{w}$ where $\overline{w}$ is the average value of the nonzero coefficients $|(w_k^j)_{exact}|$. The RMSE and NRMSE averaged over $10$ experiments are given in Table \ref{tab:examples}. The results show that the method performs well, even in the case of unstable and chaotic systems (for which the NRMSE is slightly larger). Note that the accuracy increases with the number of data points (which is low here) but a comprehensive study of this effect is beyond the scope of this paper.

\begin{table}[h]
	\centering
		\begin{tabular}{lcc}
		\hline
		   & RMSE & NRMSE \\
			\hline
a) Van der Pol & $0.035$ & $0.035$ \\
b) Unstable equilibrium & $0.157$ & $0.092$ \\
c) Chaotic Lorenz & $0.569$ & $0.074$ \\
		\hline
		\end{tabular}
		\caption{(Normalized) root mean squared error over $10$ simulations.}
		\label{tab:examples}
\end{table}

\subsection{Input and process noise}
\label{sec:example_input}

We consider the forced Duffing system
\begin{eqnarray}
\label{Duff}
\dot{x}_1 & = & x_2 \\
\dot{x}_2 & = & x_1-x_1^3-0.2 \, x_2 + 0.2 \, x_1^2 \, u
\end{eqnarray}
where $u$ is the input. With $u(t)=\cos(t)$, we obtain $51$ snapshots over $[0,10]$ (sampling period $T_s=0.2$) for $5$ trajectories with initial conditions on $[-1,1]^2$. Our identification method provides a good estimation of the vector field (including the forcing term $0.2 \, x_1^2 \, u$). The RMSE computed over all coefficients (including those related to the forcing term) and averaged over $10$ simulations is equal to $0.025$.

We now consider the effect of process noise and replace the forcing term in \eqref{Duff} by a strong white noise with $\sigma_{proc}=1$ (note that we still add measurement noise with $\sigma_{meas}=0.01$). Data points are obtained as previously, but for $10$ trajectories computed with the Euler-Maruyama scheme. The method is robust against process noise, and recovers the vector field with a RMSE (averaged over $10$ simulations) equal to $0.078$.

\subsection{Network reconstruction}

We consider a $12$-dimensional system of the form
\begin{equation}
\label{dyn_network}
\dot{x}_j = -\xi_{j} x_j  + \sum_{k=1}^3 \zeta_{j,k} \, x^{\sigma^k_{j,1}}_{\nu^k_{j,1}} x^{\sigma^k_{j,2}}_{\nu^k_{j,2}} \qquad j=1,\dots,12
\end{equation}
where the coefficients $\xi_{j}$ are chosen according to a uniform distribution on $[0,1]$ and $\zeta_{j,k}$ are distributed according to a Gaussian distribution of zero mean and standard deviation equal to one. The subscripts $\nu^k_{j,1}$ and $\nu^k_{j,2}$ are integers randomly chosen on $\{1,\dots,12\}$ and the exponents $\sigma^k_{j,1}$ and $\sigma^k_{j,2}$ are also integers randomly chosen on $\{0,1,2,3\}$ so that $\sigma^k_{j,1}+\sigma^k_{j,2}\in \{2,3\}$. The first term in \eqref{dyn_network} is a linear term that ensures local stability of the origin. The other terms are quadratic and cubic nonlinearities that define an interaction network between the different states. We say that there is a directed link from $x_l$ to $x_j$ if $\nu^k_{j,r}=l$ and $\sigma^k_{j,r} \neq 0$ for some $k \in \{1,2,3\}$ and $r \in \{1,2\}$.

\begin{figure}[h]
   \centering
    \includegraphics[width=7cm]{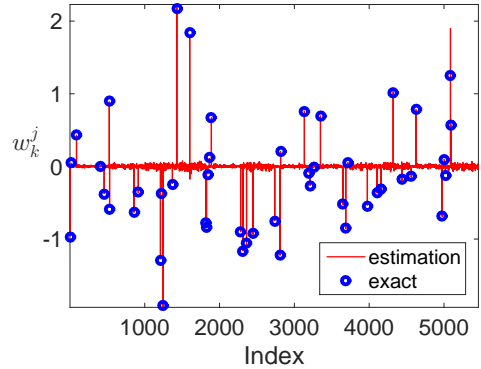}
   \caption{The method is efficient to recover all the coefficients of the polynomial vector field of a $12$-dimensional system. The exact nonzero values $w_k^j$ are depicted in blue. The estimated values are in red.}
   \label{Koopman_ident_network}
\end{figure}

For $500$ different trajectories with initial conditions on $[-1,1]^{12}$, we consider three snapshots obtained at time $t\in\{0,0.5,1\}$ (sampling period $T_s=0.5$). Figure \ref{Koopman_ident_network} shows that, with this dataset, the method provides an accurate estimation of all the coefficients of the polynomial vector field ($\textrm{RMSE}=0.021$). Note that we consider a total of $1500$ data points to identify $12 N_F=5460$ coefficients $w_k^j$. As we decrease the number of data points, results are less accurate but still acceptable ($\textrm{RMSE}=0.106$ with $400$ trajectories and $\textrm{RMSE}=0.151$ with $300$ trajectories). More advanced methods promoting sparsity (e.g. LASSO method \cite{Lasso}) could further improve the accuracy of the results.

Network reconstruction aims at predicting links in the network. Very small coefficients are mainly due to measurement noise and have an exact value equal to zero. Also we consider that there is a link from $x_l$ to $x_j$ if at least one coefficient $w_k^j$ related to a monomial containing $x_l$ is above a given threshold value. With a threshold value equal to $0.1$, we obtain a true positive rate (i.e. number of correctly identified links divided by the actual number of links) equal to $0.875$ and a false positive rate (i.e. number of incorrectly identified links divided by the actual number of missing links) equal to $0.023$. It is noticeable that, in addition to this good performance, the method also provides the nature of the links (e.g. quadratic, cubic) and their weight (i.e. values $w_k^j$).

\section{Conclusion}
\label{conclu}

We have proposed a novel method for the identification of nonlinear systems. This method is based on a lifting technique which recasts nonlinear system identification as the identification of the linear Koopman operator in an infinite-dimensional space of observables. A key advantage of the method is that it relies only on linear techniques and is an indirect method that does not require the estimation of time derivatives. It follows that the method is well-suited to model low sampling rate data sets. It is also robust to noise and efficient to recover the vector field of different classes of systems.

This paper presents preliminary results that open the door to further developments of Koopman operator lifting techniques for nonlinear system identification (e.g. use other bases to identify non-polynomial vector fields). 
Since the method is based on the (finite-dimensional) truncation of the (infinite-dimensional) Koopman operator, a theoretical study of its convergence should also be provided in future work.

\section{Acknowledgments}

The authors acknowledge support from the Luxembourg National Research Fund (FNR 14/BM/8231540).




\bibliographystyle{plain}

\end{document}